\documentclass[a4paper,superscriptaddress,twocolumn,prX,showkeys,aps]{revtex4}

\usepackage[utf8x]{inputenc}
\usepackage{amsmath}
\usepackage{amssymb}
\usepackage[english]{babel}
\usepackage[pdftex]{graphicx}
\usepackage{pst-all}

\newcommand{\ab}[1]{\frac{\mathrm{d}}{\mathrm{d} #1}}
\newcommand{\halb}{\frac{1}{2}}

\newcommand{\vx}{\vec{x}}
\renewcommand{\vr}{\vec{x}}

\newcommand{\vk}{\vec{k}}

\newcommand{\vu}{\vec{u}}

\newcommand{\eqdot}{\,.}													% Punkt am Ende von Gleichung
\newcommand{\eqcomma}{\,,}												% Komma am Ende von Gleichung
\renewcommand{\vec}[1]{\boldsymbol{#1}}

\newcommand{\dintt}{\int \! \mathrm{d}t \,}

\newcommand{\dintk}{\int \! \mathrm{d}^2k \,}
\newcommand{\dintkk}{\int \! \mathrm{d}^2k' \,}
\newcommand{\dintkt}{\int \! \mathrm{d}^2k \, \mathrm{d}t \,}

\newcommand{\dintxt}{\int \! \mathrm{d}^2x \, \mathrm{d}t \,}

\newcommand{\dintxx}{\int \! \mathrm{d}^2x' \,}

\begin{document}

\title{Gaussian Vortex Approximation to the Instanton Equations of two-dimensional Turbulence}

\author{K. Kleineberg}
\email{kolja.kleineberg@uni-muenster.de}
\affiliation{Institute for Theoretical Physics, University of Münster,
Wilhelm-Klemm-Str. 9, D-48149 Münster, Germany}
\author{R. Friedrich}
\affiliation{Institute for Theoretical Physics, University of Münster,
Wilhelm-Klemm-Str. 9, D-48149 Münster, Germany}
\date{\today}

\begin{abstract}
 We investigate two-dimensional turbulence within the Instanton formalism which determines the most probable field in a stochastic classical field theory starting from the Martin-Siggia-Rose path integral. We perform an approximate analysis of these equations based on
a variational ansatz using elliptical vortices. The result are evolution equations for the positions and the shapes of the vortices.
We solve these ordinary differential equations numerically. The extremal action for the two-point statistics is determined by the merging of two elliptical vortices.
 We discuss the relationship of this dynamical system to the inverse cascade process of two-dimensional turbulence. 
\end{abstract}


\maketitle

\section{Introduction} 
Over one and a half centuries after the discovery of the Navier-Stokes equations, the theoretical understanding of turbulent flows remains quite limited although phenomenological theories (see \cite{frisch:kolmogorov:1995}) capture the main characteristics of turbulent flows quite well. However, no generally accepted
derivation of the phenomenological theories from the basic hydrodynamical equations has been possible. The challenge consists in connecting
dynamical systems theory with nonequilibrium statistical physics \cite{falkovich:cardy}. 

The study of two-dimensional turbulence is motivated by the fact that whenever one spatial direction is significantly constrained (e.g. layers in the atmosphere or all kinds of surface dynamics) the flow can be described as quasi two-dimensional. In a two-dimensional space, the behavior of turbulent flows differs significantly from the three-dimensional case. These differences arise due to an additional conservation law for the enstrophy which leads to a new  phenomenon  called the inverse cascade discovered in 1967 by Kraichnan \cite{kraichnan:1967}. %Essentially, the formation of an inverse cascade consists of the emergence of large scale vortices  
%from initially small scale vortex structures. 
Numerically, a stationary 
inverse energy cascade can be generated by a stochastic forcing of the flow 
at small scales while dissipating energy at large scales. 

The stochastically forced vorticity equation of two-dimensional turbulence can 
be treated as a stochastic classical field theory using the Martin-Siggia-Rose 
path integral formalism \cite{msr}. Path integrals may be treated either 
perturbatively using renormalized perturbation theory or within 
the Instanton formalism by an expansion of the field around the most 
probable field configuration. This formalism has been applied by several
groups to the Burgers equation (see \cite{burgers:instanton,moriconi:2008,weinan,falkovich:burgers}),
%, \cite{moriconi:2008}, \cite{weinan} or \cite{falkovich:burgers}), 
to the problem of
turbulent advection and to estimate the tails of the velocity probability distribution of fully developed   turbulence \cite{falkovich:instanton:2011}. We also refer the readers to earlier works by Falkovich \cite{falkovich:velocity} and Moriconi \cite{moriconi:vorticity}.

The purpose of the present contribution is to investigate the dynamics of
the Instanton equations for the inverse cascade in two-dimensional turbulent flows. Thereby, we do not focus on the concrete evaluation of the extremal
action and the shape of the probability distribution. As has been pointed out
by \cite{falkovich:instanton:2011}, a reasonable estimation of the probability
distribution has to take into account fluctuations around the Instanton. 
We rather address the question whether one can identify signatures of the mechanism
underlying the inverse cascade already on the basis of the $N$-point 
Instantons. 

Our approach consists of an ansatz for the Instanton field based
on a superposition of elliptical, deformable vortices. We 
derive the most probable evolution of these vortices by extremalizing the 
Martin-Siggia-Rose \cite{msr} action (see eq. \eqref{msr-action}).
This ansatz is motivated by the observation that the Instanton equations 
in the inviscid limit and in the 
presence of small scale forcing are related to the dynamics of point vortices (see \cite{kirchhoff,helmholtz,aref:pointvortex}). 

We design a numerical procedure to solve the resulting evolution equations for the positions and the elliptical shapes explicitly for two vortices. This gives 
the two-point statistics in the Instanton approximation. Our aim here is not the complete determination of this action, which is a project for future
research. The aim of the present paper is to investigate the deterministic dynamics of the Instanton equations. Our variational ansatz using elliptical point vortices
demonstrates that the action is essentially determined by the interaction of two elliptical vortices. Thereby, the elongation of the main axis of the elliptical
vortices plays a crucial role. This elongation leads in the case of the interaction of vortices with like-signed circulation to an accelerated relative motion (see also \cite{eyink:2006}).% We discuss the relationship of the obtained dynamic to the process of the inverse cascade. 

The structure of this paper is as follows. We start with the investigation of the dynamics of Instanton point vortices which do not exhibit an inverse cascade. We present a generalized model that takes into account the deformation of vortices due to strain and calculate the interaction of two vortices numerically. We discuss the relationship of this model to the inverse cascade process.

% This combination of elongation due to mutual shear and enhanced relative acceleration is the signature of the inverse cascade in the Instanton approximation.  
% 
% It is interesting to note that the dynamics of the Instanton equations is closely related to a rotor model of the inverse cascade
% recently introduced 
% by Friedrich and Friedrich \cite{friedrich:vortex:2011}. The rotor model treats pairs of point vortices with equal circulations glued together by an inelastic spring. 
% This pair of point vortices mimics an elliptical vortex and generates a far field qualitatively similar to an elliptical vortex. Numerical and analytical calculations show
% that the dynamics of rotors is governed by the clustering of rotors of each positive and negative circulation respectively, leading to the formation of large scale
% vorticity patches. A similar dynamics is obtained in the Instanton equation.  
% Hence, the Instanton calculation supports the view that elongation and thinning of small scale vortices and the resulting clustering dynamics leads to the 
% formation of the inverse cascade in two dimensional turbulence. 

\section{Instanton equations}

We start from the two-dimensional vorticity equation for incompressible fluids with $\omega = \nabla \times \vec{u}$
\begin{equation}
 \partial_t \omega(\vec{x},t) + \vec{u}(\vec{x},t) \cdot \nabla \omega(\vec{x},t) - \nu \Delta \omega(\vec{x},t) = f(\vec{x},t)        
 \label{vorticity_eqn}
\end{equation}
where $f(\vec{x},t)$ denotes an external Gaussian white noise forcing which is specified by
\begin{equation}
\langle f(\vec{x},t)f(\vec{x}',t') \rangle =Q(\vec{x}-\vec{x}')\delta(t-t') \eqdot
\end{equation}
 The velocity field can be expressed in terms of the vorticity field via the Biot-Savart law according to
\begin{equation}
 \vec{u}(\vec{x},t) = \int \frac{\mathrm{d}^2x'}{2 \pi} \vec{e}_z \times \frac{\vec{x}-\vec{x}'}{|\vec{x}-\vec{x}|^2} \omega(\vec{x}',t) \eqdot
\end{equation}
Within the Martin-Siggia-Rose formalism \cite{msr}, all moments are derived from the partition functional which is given by the path integral
\begin{align}
\begin{split}
\mathcal{Z} & = \int \! \mathcal{D}\omega \, \mathcal{D}\hat{\omega} \, e^{\mathrm{i} S + \dintxt \left[ \eta(\vec{x},t) \omega(\vec{x},t) %+ \hat{\eta}(\vec{x},t) \hat{\omega}(\vec{x},t
) \right]}
\\ & =
\int \! \mathcal{D}\omega \, \mathcal{D} \hat{\omega} \, e^{\mathrm{i} \tilde S}
\end{split}
\end{align}
 with the extended Martin-Siggia-Rose action
\begin{multline}
\tilde S = \dintxt \hat{\omega}(\vec{x},t) \biggl[\partial_t \omega(\vec{x},t) + \vec{u}(\vec{x},t) \cdot \nabla \omega(\vec{x},t)  \\  - \nu \Delta \omega(\vec{x},t) +\frac{\mathrm{i}}{2} \dintxx  Q(\vec{x}-\vec{x}') \hat{\omega}(\vec{x'},t)\biggr]  
\\
-\mathrm{i} \dintxt  \eta(\vec{x},t) \omega(\vec{x},t) 
\eqdot
\label{msr-action}
\end{multline}

The partition functional $\mathcal{Z}$ can be related to the Laplace transform
of the $N$-point vorticity probability distribution by evaluating $\mathcal{Z}$ 
for the field $\eta(\vx,t)=\sum_{i=1}^n \alpha_i \delta(\vec{x}-\vec{x}_i)$. 

The Instanton equations are derived by performing a variation of the action 
(\ref{msr-action}) with respect to the vorticity ($\delta S / \delta \omega = 0$) and the auxiliary field ($\delta S / \delta \hat{\omega} = 0$). This procedure yields two coupled partial differential equations that describe the evolution of the most probable field configuration 
\begin{align}
\begin{split}
 \partial_t \omega  = & - \vec{u} \cdot \nabla \omega + \nu \Delta \omega - \mathrm{i} \dintxx Q(\vec{x} - \vec{x}') \hat{\omega}(\vec{x}',t) 
\\
 \partial_t \hat{\omega} = & -  \int \frac{\mathrm{d}^2x'}{2 \pi} \hat{\omega}(\vec{x}',t) \frac{\vec{e}_z \times (\vec{x}-\vec{x}')}{|\vec{x}-\vec{x}'|^2} \cdot \nabla \omega(\vec{x}',t) \\ & - \vec{u} \cdot \nabla \hat{\omega} - \nu \Delta \hat{\omega} + \eta
(\vec{x},t) \eqdot
\end{split}
\label{instantion_equations}
\end{align}
The extremal action is then given by
\begin{equation}
\mathrm{i} S_{\mathrm{extr}}=-\frac{1}{2}   
\dintxx \hat\omega(\vec{x},t)Q(\vec{x}-\vec{x}')\hat \omega({\vec x}',t) \eqdot
\end{equation}

The equation for $\omega$ has to be evaluated forward in time from 
$t^*<0$ to $0$, whereas the equation for $\hat{\omega}$ has to be solved backward in time
from $0$ to $t^*<0$. This is in accordance with the different sign of viscosity. For the auxiliary field we have the initial condition 
\begin{equation}
  \hat{\omega}(\vx,0) = \sum_{i=1}^n \alpha_i \delta(\vec{x}-\vec{x}_i)      
\label{auxiliary_initial}                                                                                                                                                    \end{equation}
which results from considering the $n$-point statistics of the field 
at spatial points $\vec{x}_i$
at time $t=0$ by setting $\eta = \delta(t) \sum_{i=1}^n \alpha_i 
\delta(\vec{x}-\vec{x}_i)$. 

\section{Zero-Viscosity Limit}

In order to motivate our further treatment, let us consider the 
evolution equation for the auxiliary field in the
limit of vanishing viscosity under the neglection of the third term on 
the left-hand side of equation (\ref{vorticity_eqn})
\begin{equation}
 \partial_t \hat{\omega} +\vec{u} \cdot \nabla \hat{\omega} 
=\sum_{i=1}^n \alpha_i \delta(\vec{x}-\vec{x}_i)\delta(t) \eqdot
\label{instantion_equations2}
\end{equation}
The solution to this equation is given by a sum over delta-functions
\begin{equation}
\hat{\omega}(\vec{x},t)=\sum_{i=1}^n \alpha_i \delta(\vec{x}-\vec{X}(\vec{x}_i,t))
\end{equation}
where $\vec{X}(\vec{x}_i,t)$ denotes the Lagrangian path (for $t\le 0$) of a passive particle moving in the velocity field
$\vec{u}(\vec{X},t)$ ending at $\vec{x}_i$ at time $t=0$ governed by the evolution equation
\begin{equation}
\frac{\partial }{\partial t} 
 \vec{X}(\vec{x},t)=\vec{u}(\vec{X}(\vec{x},t),t) 
\end{equation}
with the velocity field related to the field $\omega({\vx},t)$
by Biot-Savart's law.

The temporal evolution for the Instanton field $\omega(\vec{x},t)$ in the 
zero-viscosity limit is then given by
\begin{equation}
 \partial_t \omega + \vec{u} \cdot \nabla \omega = - \sum_{i=1}^n  Q(\vec{x} -\vec{X}(\vec{x}_i,t) ) \alpha_i  \eqdot
\end{equation}
This equation can also be solved using the Lagrangian path ${\vec{X}}({\vec{x}},t)$ by evaluating
\begin{multline}
\omega(\vec{X}(\vec{x},t))= \\-
\sum_{i=1}^n \alpha_i
\int_{t^*}^t \! \mathrm{d}t' \,  Q(\vec{X}(\vec{x},t') -\vec{X}(\vec{x}_i,t') )  \eqdot
\end{multline}
For small-scale stirring we can approximate the correlation $Q(\vec{r})$ by
a $\delta$-function.

Finally, we have to determine the velocity field, which is obtained by a 
combination of Biot-Savart's law and the representation of the vorticity field
\begin{equation}
\vu(\vx,t) =
\sum_{i=1}^n \alpha_i \left[\int_{t^*}^t \! \mathrm{d}t' \, q
\vec{e}_z \times 
\frac{\vx-\vec{X}(\vx_i,t')}{2\pi |\vx-\vec{X}(\vx_i,t')|^2} \right] \eqdot 
\end{equation}
For the sake of simplicity, we approximated $Q(\vec{x}-\vec{x}')$ by a
delta function, $q\delta(\vec{x}-\vec{x}')$. We obtain 
an extension of the usual point vortex dynamics by the fact that the 
circulations is time dependent $ \propto (t-t^*) \alpha_i q$.   

This dependence of the initial time $t^*$ can be taken into account by a time shift
$\tau=t-t^*$. The Instanton point vortex dynamics is given by ($\vx_i(\tau)=
\vec{X}(\vx_i,\tau + t^*)$)
\begin{equation}
\ab{\tau} \vx_i=\tau \sum_{j=1}^n
\vec{e}_z \times 
\frac{\vx-\vx_j(\tau)}{2\pi |\vx-\vx_j(\tau)|^2} \eqdot
\end{equation}
The transformation $\tau'=\halb \tau^2$ yields the usual point vortex dynamics.
Therefore, we conclude that the inviscid Instanton point vortex dynamics
is similar to the usual point vortex dynamics in an accelerated timeframe which consists of replacing the $\tau$-dependence
by a $\tau^2$ dependence.  

On the basis of the point vortex approximation, there is no mechanism which
could be related to the inverse cascade.

\section{Deformable Vortices}

In this section we sketch a generalization taking into account the finite 
extension of the vortices as well as a deformation of the vortices due to
shear generated by distant vortices. This will motivate our
approximate treatment of the Instanton equations. 

We start with the auxiliary field, which for $t\rightarrow 0$ actually has
to tend to a superposition of $\delta$-functions, i.e. to a superposition 
of point vortices. For negative times, viscosity will broaden the vortices
whereas presence of the advective term and the type of vortex stretching term
will lead to a deformation of the circular shape. Thus, the straightforward
extension is an ansatz in terms of elliptical point vortices
\begin{equation}
\hat \omega(\vx,t)=\sum_{j=1}^m \frac{\hat \Gamma_j(t)}
{2\pi\sqrt{\det \hat{C}_j}}e^{-\frac{1}{2}(\vx-\hat\vx_j(t))
\hat{C}_j^{(-1)}(\vx-\hat\vx_j(t))} \eqdot
\end{equation}
From the condition (\ref{auxiliary_initial}) we conclude that 
\begin{equation}
\hat\Gamma_j(0)=\alpha_j,  
\hat{C}_j(0)=0, \text{ and }  \hat\vx_j(0)=\vx_j \eqdot
\end{equation}

A similar ansatz is performed for the Instanton field $\omega(\vx,t)$
\begin{multline}
\omega(\vx,t)= \\ \sum_{j=1}^n \frac{\Gamma_j(t)}
{2\pi\sqrt{\det {C}_j(t)}}e^{-\frac{1}{2}(\vx-\vx_j(t))
{C}_j(t)^{(-1)}(\vx-\vx_j(t))} \eqdot
\end{multline}
The initial condition for $t^\ast 
\rightarrow -\infty $, $\omega(\vx,t)=0$ is
defined by $\Gamma_j(t^\ast)=0$, $C_j(t^\ast)$ and the
initial locations $\vx_j(t^\ast)$. The initial condition 
$\omega(\vx,t^\ast)=0$ can be realized by $\Gamma_j(t^\ast)=0$.

It is evident that we could extend our ansatz by including, besides the elliptical ansatz, higher-order terms.
%, e.g. by an expansion using Hermite polynomials,
%as has recently been suggested by Wayne \cite{Wayne}.  
Such an expansion with an additional averaging over the higher-order 
contributions is needed in order to go beyond the bare Instanton approximation.
This program would lead to an effective action in much the same way
as has been demonstrated by Falkovich et al. \cite{falkovich:instanton:2011} in 
their treatment of the statistics of the vorticity increment 
of the direct cascade. They separated the fields into an axisymmetric 
and a nonaxisymmetric part, where the nonaxisymmetric fluctuations where
integrated out to obtain an effective action for the axisymmetric part, which,
in turn was treated in the Instanton approximation. 
We conjecture that such an effective action could be characterized by a 
modified noise correlation $Q(\vx-\vx')$ as well as a modified 
Biot-Savart law,
$\vu(\vx,t)=\vec{e}_z \times \nabla 
F(-\Delta) \omega(\vx,t)$, where $F(-\Delta)$ is a function of 
the Laplacian.     

In the present
article we will be content with the elliptic vortex approximation.
Our main point here is that the behavior of the elliptical vortices
depends on the signs of $\Gamma_j$ and, hence, on the sign of $\alpha_j$.
We will explicitly demonstrate that two 
like-signed vortices will attract each other due to this deformation. 

\section{Action for elliptic vortices}

In this section we consider the Martin-Siggia-Rose action for a field which
consists of a superposition of $N$ elliptical vortices. Each vortex is 
determined by its location $\vec{x}_i(t)$, its amplitude $\Gamma_i(t)$
 and a matrix $C_i(t)$ determining
the orientation and the semi-axis of the elliptical vortices. 
Similar quantities are introduced for the auxiliary field.
By introducing
these collective variables, the Instanton equations, which 
are partial differential equations, are approximated by ordinary
differential equations, which are the Euler-Lagrange equations for the
MSR-action, written in collective variables. These equations turn out to
be extensions of the inviscid point vortex Instantons. The inclusion of
the elliptical shape of the vortices, however, will lead to a modified
vortex dynamics.

As a starting point we consider the Martin-Siggia-Rose action 
(eq. \eqref{msr-action}) in Fourier space
\begin{multline}
 S = \dintkt \hat{\omega}(-\vec{k},t) \biggl\{ \partial_t \omega(\vec{k},t) + \nu k^2 \omega(\vec{k},t)
\\
 - \mathrm{i} \vec{k} \cdot \dintk' \vec{u}(\vec{k}') \omega(\vec{k}-\vec{k}',t) \omega(\vec{k}',t)  \biggr\}
\\
+ \frac{\mathrm{i}}{2} \dintkt \hat{\omega}(-\vec{k},t) Q(\vec{k}) \hat{\omega}(\vec{k},t)
\label{action_fourier}
\end{multline}
where we defined $\vec{u}(\vec{k}) = \mathrm{i} \frac{\vec{e}_z \times \vec{k}}{4 \pi^2 k^2}$. 
We make an elliptic ansatz for the vorticity and the auxiliary field according to \begin{align}
 \begin{split}
 \omega(\vec{k},t) & = \sum_{i=1}^n \Gamma_i e^{\mathrm{i} \vec{k} \cdot \vec{x}_i - \frac{1}{2} \vec{k} C_i \vec{k}} \equiv \sum_{i=1}^n \Gamma_i \omega_i(\vec{k},t) \\
  \hat{\omega}(\vec{k},t) &  = \mathrm{i} \sum_{i=1}^m \hat{\Gamma}_i e^{i \vec{k} \cdot \vec{\hat{x}}_i - \frac{1}{2} \vec{k} \hat{C_i }\vec{k}} \equiv \mathrm{i} \sum_{i=1}^m \hat{\Gamma}_i \hat{\omega}_i(\vec{k},t) \eqdot
\end{split}
\label{ellip_ansatz}
\end{align}
$C_i$ and $\hat{C}_i$ are $2\times2$ symmetric matrices and $\Gamma_i$ and $\hat{\Gamma}_i$ are the circulations of each vortex or elliptic structure in the auxiliary field.  

Here, we assume that the quantities $\Gamma_i(t)$, $\vec{x}_i(t)$, $C_i(t)$ as
well as the corresponding quantities for the auxiliary field depend on time.
Extremalization of the MSR-action will lead us to an approximation 
of the Instanton equations in terms of a coupled set of
ordinary differential equations. As we have discussed in the previous section,
we include the deformation of the vortices since this deformation seems to play
a major role for the energy transport in the inverse cascade.

We have to specify the boundary conditions for the fields with respect to
time. The conditions for the auxiliary field are easily formulated. From
the representation (\ref{auxiliary_initial}) we obtain
\begin{equation}
\hat \vx_i(0)=\vx_i \text{ and } \hat C_i(0)=0 \eqdot
\end{equation}
The initial condition for the field $\omega(\vec{x},t)$ has to be formulated 
for $t\rightarrow -\infty$, in particular we have $\omega(\vec{x},t\rightarrow -\infty)=0$.
As we will see below, this can be achieved by putting $C_i(-\infty)=0$, i.e. 
a infinitely extended vortex. Another possibility would be to set 
$\Gamma_i(t^\ast)=0$ for a large negative time $t^\ast$. The reason will become
more evident below, when we formulate the Euler-Lagrange equation corresponding
to the MSR-action for the elliptical vortices.
 
We plug the ansatz into the action \eqref{action_fourier} and integrate 
the action with respect to $\vk$ which yields
\begin{multline}
 S = \sum_{i=1}^m \sum_{j=1}^n  \dintt \hat{\Gamma}_i \Gamma_j
\Biggl\{ \frac{\dot{\Gamma}_j}{\Gamma_j} - \dot{\vx}_j \cdot \hat{\nabla}_i 
+ \halb \hat{\nabla}_i \dot{C}_j \hat{\nabla}_i 
\\
- \mathrm{i} \sum_{k=1}^n \Gamma_k A_{jk}(\hat{\nabla}_i) 
 - \nu \hat{\nabla}_i \mathcal{I} \hat{\nabla}_i
 + \frac{\mathrm{i}}{2} \frac{\hat{\Gamma}_j}{\Gamma_j} K_j(\hat{\nabla}_i) \Biggr\} W_{ij} \eqcomma
\label{wirkung_ellipsen}
\end{multline}
where $\hat{\nabla}_i$ denotes $\nabla_{\hat{\vx}_i}$ and we defined
\begin{align}
\begin{split}
 A_{jk}(\hat{\nabla}_i) & = \mathrm{i} \dintkk e^{(\vx_j-\vx_k) \cdot \hat{\nabla}_i 
+ \halb \hat{\nabla}_i (C_j+C_k) \hat{\nabla}_i}
\\ & \qquad \times e^{ \frac{i}{2} \hat{\nabla}_i C_j \vk' 
+ \frac{\mathrm{i}}{2} \vk' C_j \hat{\nabla}_i} \vu(\mathrm{i} \hat{\nabla}_i) \cdot \hat{\nabla}_i  \eqcomma 
\\
K_j(\hat{\nabla}_i) & = \mathrm{i} Q(\mathrm{i} \hat{\nabla}_i) e^{-(\hat{\vx}_j-\vx_j) \cdot \hat{\nabla}_i + \halb \hat{\nabla}_i (\hat{C}_j-C_j) \hat{\nabla}_i}  \eqcomma 
\\
W_{ij} & = \mathrm{i} \frac{2 \pi}{\sqrt{\det[\hat{C}_i + C_j]}} 
\\ & \times e^{-\halb (\hat{\vx}_i-\vx_j) (\hat{C}_i+C_j)^{-1} (\hat{\vx}_i-\vx_j)} \eqdot
\end{split}
\label{def_a_k}
\end{align} 
Equation \eqref{wirkung_ellipsen} is the action for elliptic vortices and elliptic structures in the auxiliary field according to the ansatz from eq. \eqref{ellip_ansatz}. We will derive the evolution equations for the parameters in the ansatz by applying a variation according to the Euler-Lagrange equations. To this end, we make the following assumptions.

\begin{enumerate}
 \item We assume that we have the same number $n=m=N$ of elliptic vortices and elliptic structures in the auxiliary field. 
 \item The elliptic structures in the auxiliary field are positioned such that the $i$-th vortex is always close to the $i$-th elliptic structure, which means
\begin{equation}
 \left| \vx_i(t) - \hat{\vx}_i(t) \right| \ll \text{Size of vortices} \eqdot
\label{annahme_2}
\end{equation}
\item The vortices and elliptic structures are isolated, hence 
their overlap can be omitted. This yields $ \forall i \neq j$
\begin{subequations}
\begin{equation}
  \left|\vx_i(t) - \vx_j(t) \right| \gg \text{Size of vortices}
\end{equation}
and $\forall i \neq j$
\begin{equation}
   \left|\hat{\vx}_i(t) - \hat{\vx}_j(t) \right| \gg \text{Size of elliptic structures} \eqdot
\end{equation}
\end{subequations}
\end{enumerate}

Hence, we only account for the diagonal terms in eq. \eqref{wirkung_ellipsen}. 
Based on this approximation we pursue the investigation of the action 
\begin{multline}
 S = \sum_{i=1}^N \dintt \hat{\Gamma}_i \Gamma_i \Biggl\{ \frac{\dot{\Gamma}_i}{\Gamma_i} - \dot{\vx}_i \cdot \hat{\nabla} + \halb \hat{\nabla} \dot{C}_i \hat{\nabla} 
\\
- \mathrm{i} \sum_{j=1}^N \Gamma_j A_{ij}(\hat{\nabla}) 
 - \nu \hat{\nabla} \mathcal{I} \hat{\nabla}
 + \frac{\mathrm{i}}{2} \frac{\hat{\Gamma}_i}{\Gamma_i} K_i(\hat{\nabla}) \Biggr\} W_{ii} \eqcomma
\label{wirkung_ellipsen_d}
\end{multline}
where we have set $i=j$ and then renamed $k$ to $j$.

\section{Evolution equations for collective coordinates}
We formulate the evolution equations for our Instanton model of elliptical vortices and the corresponding auxiliary field 
\begin{align}
\begin{split}
 \dot{\Gamma}_i &=  q \hat{\Gamma}_i \eqcomma \\
\dot{\vec{x}}_i & =  \sum_{j=1}^N \Gamma_j \vec{U}_{ij}(\vec{x}_i-\vec{x}_j)+ q \frac{\hat{\Gamma}_i}{\Gamma_i} (\vec{\hat{x}}_i - \vec{x}_i(t)) \eqcomma
\\
\dot{C}_i & =  \sum_{j=1}^N \Gamma_j \left[S_{ij}(\vec{x}_i-\vec{x}_j) C_i + C_i S_{ij}^{T} (\vec{x}_i - \vec{x}_j) \right] \\ &  + q \frac{\hat{\Gamma}_i}{\Gamma_i} \left(\Tilde{Q} + \hat{C}_i - C_i\right) + 2 \nu \mathcal{I} \eqcomma \\
  \dot{\hat{\Gamma}}_i  & =  0 \eqcomma \\
  \dot{\hat{\vec{x}}}_i & =   \sum_{j=1}^N \Gamma_j \vec{\hat{U}}_{ij}(\hat{\vec{x}}_i - \vec{x}_j) \eqcomma \\
\dot{\hat{C}}_i & =   \sum_{j=1}^N \Gamma_j \left[ \hat{S}_{ij}\left(\hat{\vec{x}}_i - \vec{x}_j\right) \hat{C}_i + \hat{C}_i \hat{S}_{ij}^{T}\left(\hat{\vec{x}}_i - \vec{x}_j\right) \right] \\ & - 2 \nu \mathcal{I} 
\label{eqnevo}
\end{split}
\end{align}
in collective coordinates as the main result of our approach. 
Note that we accounted for the previously mentioned assumptions in the derivation of these equations which can be found in the appendix \ref{evolution_elliptic}. 

These equations allow for a subset of solutions for which the locations of the
vortices of both fields $\vx_i(t)$ and $\hat \vx_i(t)$ coincide. We will discuss this case later.

\section{Two-point Instanton}

\begin{figure*}[t]
\center
\includegraphics[width=0.24\textwidth]{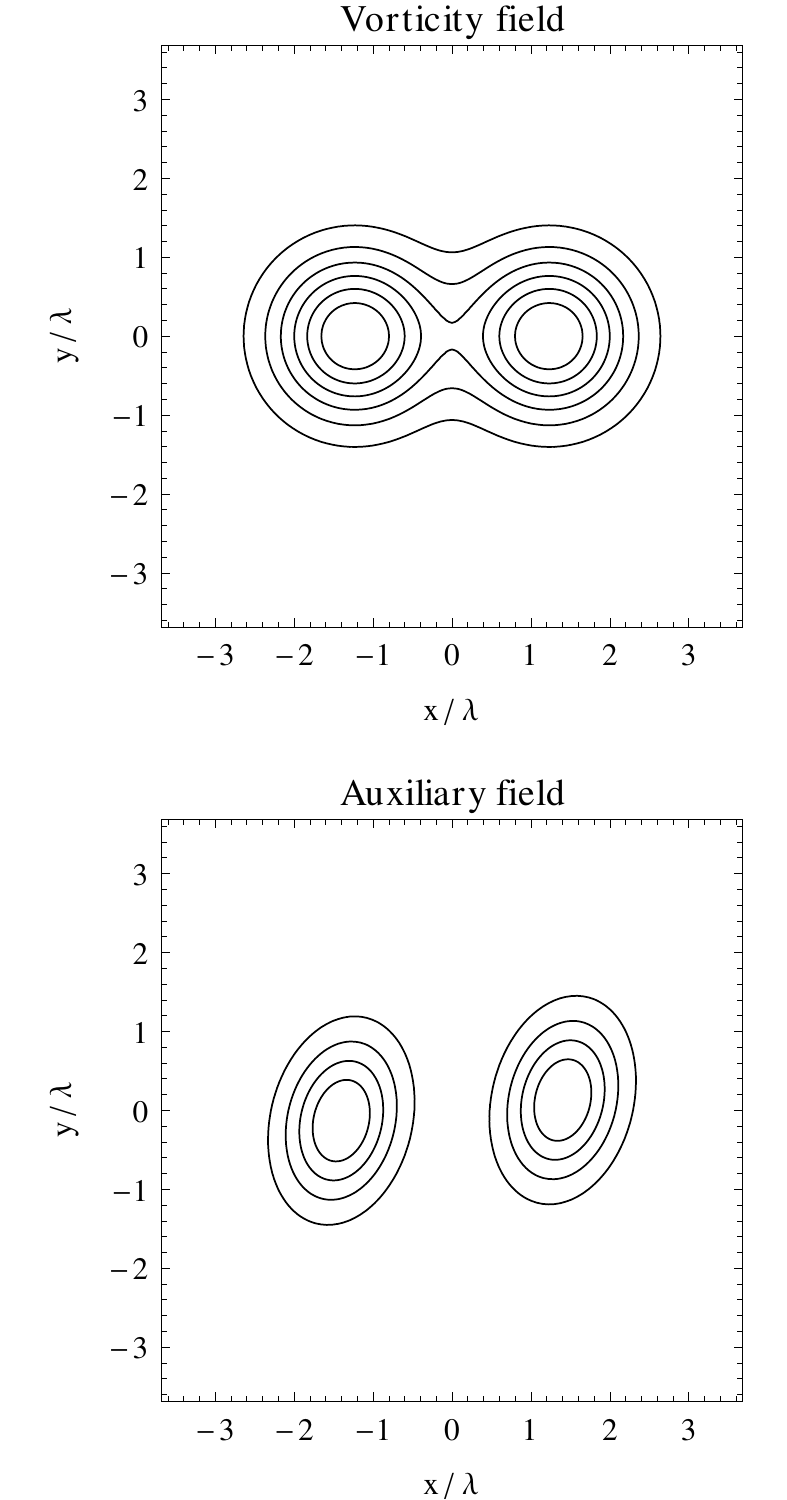}
\includegraphics[width=0.24\textwidth]{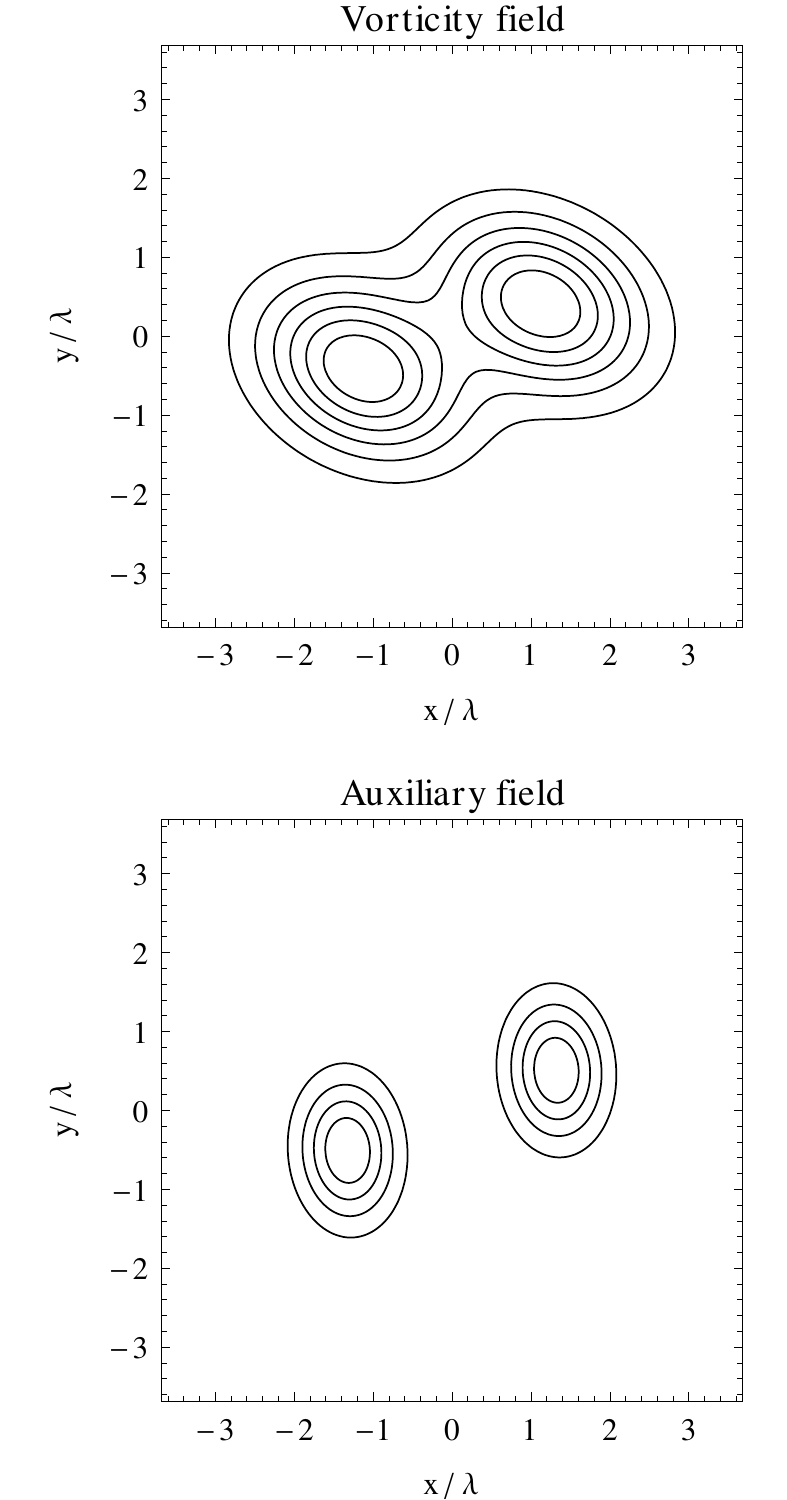}
\includegraphics[width=0.24\textwidth]{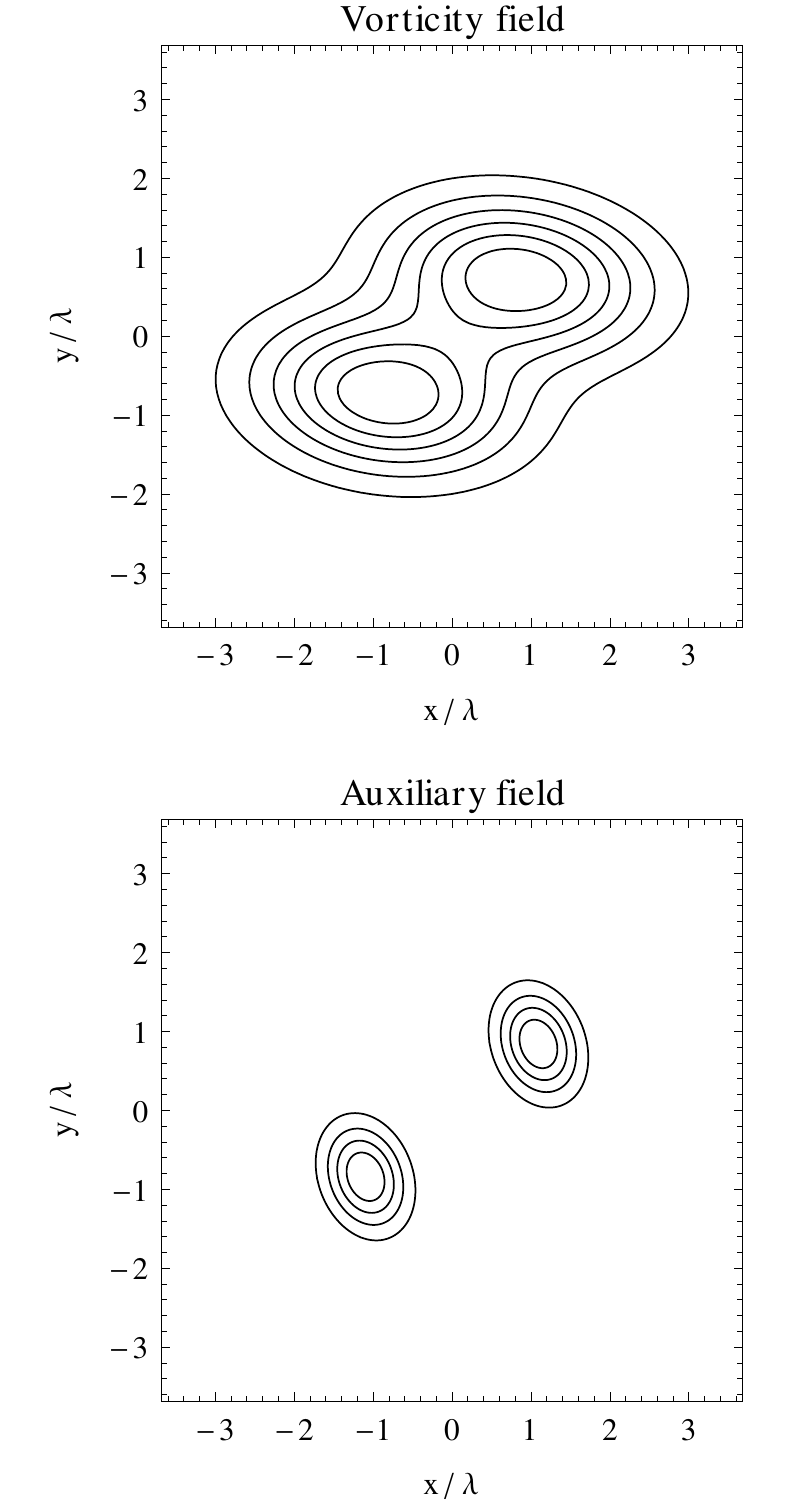}
\includegraphics[width=0.24\textwidth]{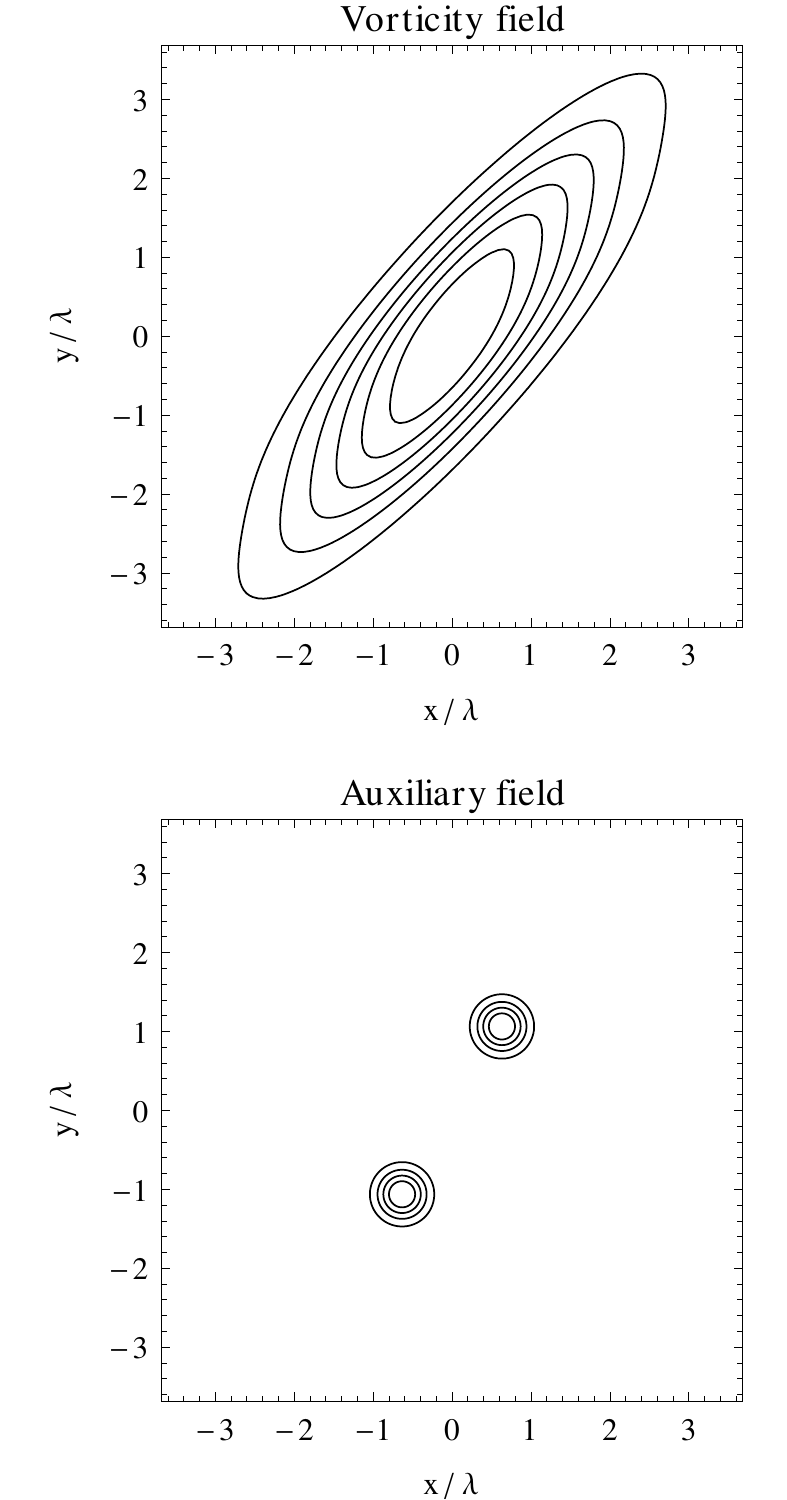}
\caption{Evolution of two vortices of same-signed circulation and their corresponding auxiliary fields. The vortices main axes become elongated and the vortices approach each other. Time increases from left to right.}
\label{elongation1}
\end{figure*}

We solve the evolution equations explicitly for two vortices with same-signed circulations. In accordance with the assumptions made earlier, we allow for two elliptically shaped structures in the auxiliary field which have to tend to infinitely small objects once time $t$ tends to $0$. This means that at time $0$ the auxiliary field consists of a superposition of two delta functions (see eq. \eqref{auxiliary_initial}) and hence is connected to the two-point statistics of the vorticity field. We call the corresponding Instanton the two-point Instanton which provides an insight to the interaction of two vortices.

We want to derive the dynamics of the two-point Instanton to understand the underlying vortex dynamics. To this end, we expand the advection and deformation and make use of an iterative calculation scheme to obtain the solution for the vorticity and auxiliary field. %The exact procedure can be found in the appendix. 
%We observe the elongation of elliptic vortices (see fig. \ref{elongation1}) and the tendency of the elongated vortices to approach each other (see fig. \ref{distplot}).

Let us consider the calculations in some more detail.
We want to calculate the evolution of two elliptic vortices and two elliptic structures governed by the evolution equations \eqref{eqnevo}. 

To this end, we expand the advection for $i \neq j$\footnote{Within the expansion of the advection and deformation for both fields, we exclude the $i = j$ term in analogy to the treatment of point vortices.}
\begin{equation}
 \vec{U}_{ij}\left(\vec{r}\right) \approx  \left[1 + \frac{1}{2} \nabla_r^T \left(C_i + C_j\right) \nabla_r\right] \vec{e}_z \times \frac{\vec{r}}{2 \pi |\vec{r}|^2} 
\label{approx_form_1}
\end{equation}
until first order in $C_i + C_j$ and define 
\begin{equation}
  \vec{V}(\vr) = \frac{1}{2 \pi r^2} \begin{pmatrix}
                             -y \\ x
                            \end{pmatrix} \eqdot
 \end{equation}
We can write $\vec{U}_{ij}(\vec{r})$ with $C^{(ij)}_{\mu \nu} = \frac{1}{2} ( C^{(i)}_{\mu \nu} + C^{(j)}_{\mu \nu} )$ as
\begin{equation}
 \vec{U}_{ij}(\vec{r}) = \vec{V}(\vec{r})  + \sum_{\alpha,\beta = 1}^2 C_{\alpha \beta}^{(ij)} \frac{\partial}{\partial x_\alpha} \frac{\partial}{\partial x_\beta} \vec{V}(\vec{r})
 %+ C^{(ij)}_{11} \vec{V}_{x,x} + C^{(ij)}_{12} \vec{V}_{x,y} \\ + C^{(ij)}_{21} \vec{V}_{y,x}  + C^{(ij)}_{22} \vec{V}_{y,y}
\label{vector_u}
\end{equation}
where $C_{\mu \nu}^{(i)}$ denotes the component $\mu \nu$ of the matrix $C_i$.

For the deformation we have to evaluate
\begin{equation}
  S_{ij}(\vec{r}) =  \left[1 + \frac{1}{2} \nabla_r^T \left(C_i + C_j\right) \nabla_r\right] \vec{e}_z \times \frac{\vec{r}}{2 \pi |\vec{r}|^2} \overleftarrow{\nabla}_r^T
\end{equation}
which we can write as
\begin{equation}
   S_{ij}(\vec{r}) =  M(\vec{r}) + \sum_{\alpha,\beta = 1}^2 C_{\alpha \beta}^{(ij)} \frac{\partial}{\partial x_\alpha} \frac{\partial}{\partial x_\beta} M(\vec{r})
%+ C^{(ij)}_{11} M_{x,x} + C^{(ij)}_{12} M_{x,y} \\ + C^{(ij)}_{21} M_{y,x}  + C^{(ij)}_{22} M_{y,y}
\label{matrix_s}
\end{equation}
with the matrix
\begin{equation}
 M(\vr) = \frac{1}{2 \pi r^4} \begin{pmatrix} 2 x y & 2y^2-r^2 \\ -2 x^2 + r^2 & -2 x y \end{pmatrix} \eqdot
\end{equation}
An analogue calculation for the auxiliary field yields
\begin{equation}
 \hat{\vec{U}}_{ij}(\vec{r}) = \vec{V}(\vec{r})  + \sum_{\alpha,\beta = 1}^2 \tilde{C}_{\alpha \beta}^{(ij)} \frac{\partial}{\partial x_\alpha} \frac{\partial}{\partial x_\beta} \vec{V}(\vec{r})
%+ \Tilde{C}^{(ij)}_{11} \vec{V}_{x,x} + \Tilde{C}^{(ij)}_{12} \vec{V}_{x,y} \\ + \Tilde{C}^{(ij)}_{21} \vec{V}_{y,x}  + \Tilde{C}^{(ij)}_{22} \vec{V}_{y,y}
\end{equation}
and
\begin{equation}
  \hat{S}_{ij}(\vec{r}) =  M(\vec{r}) + \sum_{\alpha,\beta = 1}^2 \tilde{C}_{\alpha \beta}^{(ij)} \frac{\partial}{\partial x_\alpha} \frac{\partial}{\partial x_\beta} M(\vec{r})
%+ \Tilde{C}^{(ij)}_{11} M_{x,x} + \Tilde{C}^{(ij)}_{12} M_{x,y} \\ + \Tilde{C}^{(ij)}_{21} M_{y,x} + \Tilde{C}^{(ij)}_{22} M_{y,y} 
\end{equation}
where $\Tilde{C}^{(ij)}_{\mu \nu}$ denotes $\hat{C}^{(i)}_{\mu \nu} + C^{(j)}_{\mu \nu}$ and $ \vec{r} \rightarrow \hat{\vec{x}}_i - \vec{x}_j$. \\
The auxiliary field satisfies the initial condition \hbox{$\hat{\omega}(\vec{x},0) = \sum_{i=1}^2 \alpha_i \delta(\vec{x}-\vec{\hat{x}}_i)$} according to eq. \eqref{auxiliary_initial} which corresponds to two point vortices. The vorticity field is fixed at a time $t^*<0$. Because the initial conditions for the two fields have to be fixed at different times, a straight forward calculation is not feasible. 

We solve the equations for two vortices and two elliptic structures in the auxiliary field using an iterative technique. We start from the well known solution for two point vortices (this emerges in the limit of vanishing $C_i$ and $\hat{C}_i$) located at $\vec{x}_1$ and $\vec{x}_2$ or $\hat{\vec{x}}_1$ and $\hat{\vec{x}}_2$ respectively with $\vec{x}_i(t) = \hat{\vec{x}}_i(t)$ for $i = {1,2}$. The initial condition for the auxiliary field consists of two point vortices located at $\hat{\vec{x}}_1$ and $\hat{\vec{x}}_2$ so that one can argue that $\hat{\omega}$ will behave similar to the motion of a two point vortex system. Now we calculate the evolution of the vorticity field $\omega$ according to the set of equations \eqref{instanton1} in the limit $\hat{C}_i \rightarrow 0$. Hence, we have $\hat{\vec{x}}_i(t) = U(\vartheta t) \vec{x}_i(0)$ where $U$ denotes the two-dimensional rotation matrix $SO(2)$ and $\vartheta = (\Gamma_1+\Gamma_2)/(2 \pi d^2)$. We use the approximations for the advection $\vec{U}_{ij}
$ and deformation $S_{ij}$ from equations \eqref{vector_u} and \eqref{matrix_s} to calculate $\vec{x}_i(t)$ and $C_i(t)$. The ordinary differential equations can be solved with an explicit Runge-Kutta method. 

Now we can calculate the auxiliary field $\hat{\omega}$ for $\hat{C}_i \neq 0$ if we plug in the previously calculated solution for the vorticity field given by $\Gamma_i(t)$, $\vec{x}(t)$ and $C_i(t)$. Because of $\partial_t \hat{\Gamma}_i \approx 0$ the circulation $\Gamma_i(t)$ grows linearly in time according to $q \hat{\Gamma}_i (t-t^*) + \Gamma_i(t^*)$. With the obtained solution for the elliptic structures in the auxiliary field we can calculate a more precise solution of the vorticity field and successively repeat these steps. Due to the smallness of the coupling parameter $q$ compared to the the circulations ($q \approx 0.05$ where we chose the circulations in the order of unity) this iterative procedure converges quickly. 

\begin{figure}[t]
 \center
\includegraphics[width=0.45\textwidth]{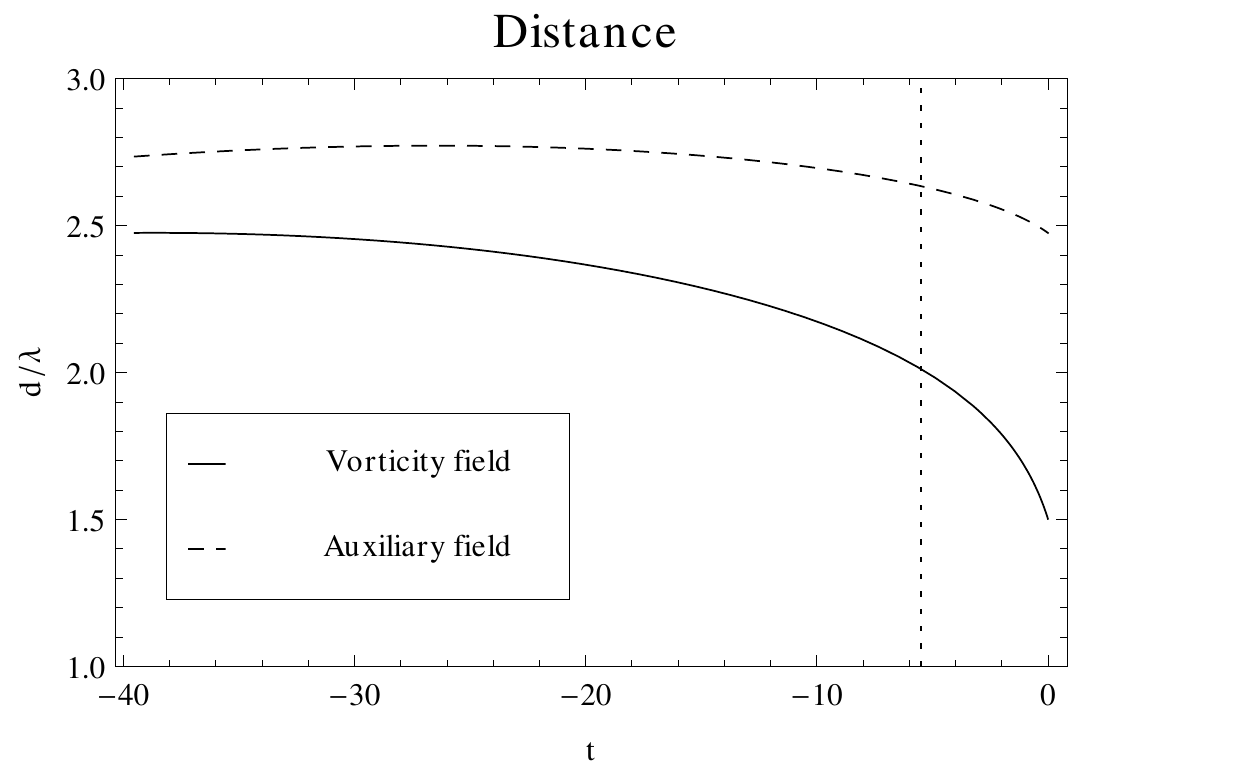}
\caption{Evolution of the distance between the centers of the vortices (solid line) and the centers of the elliptic structures in the auxiliary field (dashed line). One sees the tendency of the vortices to approach each other rapidly once the elongation becomes dominant (vertical dotted line).}
\label{distplot}
\end{figure}

The result for same-signed vortices is shown in figure \ref{elongation1}. Initially circular shaped vortices are deformed with time, forming long elliptic structures. These structures then approach each other which is shown in figure \ref{distplot}, where we plotted the distance between the centers of the vortices.

\section{Reduced equations}

We want to discuss the relationship of the elliptic model described by the set of evolution equations \eqref{eqnevo} to the inverse cascade in more detail. To this end, we formulate a reduced version of the evolution equations. We perform the following approximations:
\begin{enumerate}
 \item The position of the center of each vortex is equal to the center of each structure in the auxiliary field at each time $t$ between $t^*$ and $0$. This means
 \begin{equation}
  \vx_i = \hat{\vx}_i \qquad \forall t \in [t^*,0]
 \end{equation}
 and hence the term $q \frac{\hat{\Gamma}_i}{\Gamma_i} (\vec{\hat{x}}_i - \vec{x}_i(t))$ in \eqref{eqnevo} vanishes.
\item The structures in the auxiliary field are infinitely small and can be described by $\delta$ functions. In terms of the collective coordinates, this means
\begin{equation}
 \hat{C}_i = 0 \qquad \forall t \in [t^*,0] \eqdot
\end{equation}
As a consequence, the coupling term in the evolution equation for $C_i$ reduces to
\begin{equation}
  q \frac{\hat{\Gamma}_i}{\Gamma_i} (\Tilde{Q} + \hat{C}_i - C_i)  \rightarrow q \frac{\hat{\Gamma}_i}{\Gamma_i} \left(\Tilde{Q} - C_i\right)
\end{equation}
and hence the evolution of $C_i$ becomes decoupled.
\end{enumerate}
We obtain a reduced set of evolution equations for the vorticity field that implicitly contains the evolution of the auxiliary field given the above approximations. One can think of it as an enslavement process where the evolution of the point vortex like auxiliary field follows instantaneously the centers of the elliptic vortices. It shall also be noted that the approximation $\hat{C}_i(t) = 0$ is consistent with the initial condition at time $t=0$. The resulting equations for the vorticity field are
\begin{align}
 \begin{split}
  \dot{\Gamma}_i & = q \hat{\Gamma}_i \eqcomma\\
\dot{\vec{x}}_i & =  \sum_{j=1}^N \Gamma_j \vec{U}_{ij}(\vec{x}_i-\vec{x}_j) \eqcomma \\
\dot{C}_i & =  \sum_{j=1}^N \Gamma_j \left[S_{ij}(\vec{x}_i-\vec{x}_j) C_i + C_i S_{ij}^{T} (\vec{x}_i - \vec{x}_j) \right] \\ &  + q \frac{\hat{\Gamma}_i}{\Gamma_i} \left(\Tilde{Q} - C_i\right) + 2 \nu \mathcal{I} \eqdot
\label{eqnsimple}
 \end{split}
\end{align}
Given the solution to \eqref{eqnsimple} the evolution of the auxiliary field is well known. The circulation $\hat{\Gamma}_i$ is constant, the position $\hat{\vx}_i$ is equal to $\vx_i$ and the matrix $\hat{C}_i$ is zero. Nevertheless, the effect of the auxiliary field is still present and taken into account by the term $q \frac{\hat{\Gamma}_i}{\Gamma_i} (\Tilde{Q} - C_i)$.

It is interesting to mention that \eqref{eqnsimple} is equivalent to the equations for elliptic vortices presented by Friedrich and Friedrich in \cite{friedrich:vortex:2011} from which they derived the rotor model. The term $q \frac{\hat{\Gamma}_i}{\Gamma_i} (\Tilde{Q} - C_i)$ can be considered an overdamped spring that forces the shape of the elliptic vortices to a certain shape. In the limit of strong elongation the elliptic vortices can be approximated by pairs of point vortices connected via an overdamped spring with the equilibrium distance $D_0$ and an appropriate spring constant $\gamma/2$. This leads to the rotor model which exhibits an inverse cascade by cluster formation of same-signed vortex pairs starting from arbitrary initial conditions. See \cite{friedrich:vortex:2011} for a detailed discussion.

In summary, the effect of the auxiliary field yields the extension of the point vortex model that is responsible for breaking the $\Gamma_i \rightarrow -\Gamma_i, t \rightarrow -t$ symmetry giving rise to an inverse cascade as demonstrated by the rotor model.

\section{Discussion}

We investigated the dynamics of the Instanton equations for the inverse cascade in two-dimensional turbulent flows. Because the dynamics of point vortices as considered in the limit of vanishing viscosity do not lead to an inverse cascade we decided to introduce a more sophisticated model taking into account a finite spatial extension of the vortices as well as their ability to perform deformations on the basis of shear generated by distant vortices. 

To this end, we derived the most probable evolution of elliptic vortices by extremalizing the Martin-Siggia-Rose action under certain assumptions. The main result is given by the evolution equations in collective coordinates \eqref{eqnevo}.

We focused on the interaction of two same-signed elliptic vortices where we observed the elongation of the main axis which led to an acceleration of the relative motion. In our opinion, the combination of the elongation and the acceleration of the relative motion which leads to a reduction of the distance between two same-signed vortices is the signature of the inverse cascade. 
A reduced set of equations for the evolution of the vorticity field was derived which leads directly to the rotor model \cite{friedrich:vortex:2011}. It exhibits an inverse cascade by cluster formation of same-signed rotors which correspond to infinitely elongated elliptic vortices.

In summary, we investigated the two-point Instanton by evaluating the interaction of two deformable same-signed vortices.
%and made the connection to the rotor model which exhibits an inverse cascade. 
We identified the underlying dynamics in accordance with \cite{chen:thinning:2006} and \cite{melander:vortex} as elongation and thinning of elliptical shaped vortices due to the shear and hence the acceleration of the relative motion which leads to enhanced transport of energy and enstrophy. We conclude, that the Instanton
calculation supports the view that elongation and
thinning of small scale vortices and the resulting
clustering dynamics lead to the formation of the
inverse cascade in two-dimensional turbulence. 

It remains an interesting task for the future to generalize our approach by modifying the ansatz. One could think of vortices with deformable main axes so that one would expect to obtain S-shaped structures or an extension to the exponential of the form $\omega \propto \exp[(\vx C^{-1} \vx)^\alpha]$. The evaluation of the two-point statistics within the elliptical model poses an interesting challenge which consists of the evaluation of the transition probability which we sketched in the appendix (see eq. \eqref{transition_probability}). To this end, one has to consider solutions to the Instanton equations that deviate from the point vortex vicinity and thus disobey the assumptions we made for our approach.

% An evaluation of equations \eqref{eqnevo} for $N$ vortices as well as a generalization to three dimensions is a task for future work. One could also think about modifying the ansatz according to $\omega \propto e^{- (\vx C_i^{-1} \vx)^\alpha}$ and focus on the spectral properties of the elliptic vortex model.

% A brief introduction to the evaluation of the transition probability can be found in the appendix \ref{section_transition_probability}. The evaluation of equation \eqref{transition_probability} remains an open task for future work. We hope that the investigation of this expression will provide further insight to the statistical interpretation of the Instanton equations for two-dimensional turbulence. 

\appendix

\section{Evolution equations for elliptic vortices}
\label{evolution_elliptic}

We start from the Lagrangian
\begin{multline}
 \mathcal{L} = \sum_{i=1}^N \hat{\Gamma}_i \Gamma_i \Biggl\{ \frac{\dot{\Gamma}_i}{\Gamma_i} - \dot{\vx}_i \cdot \hat{\nabla} + \halb \hat{\nabla} \dot{C}_i \hat{\nabla} 
\\
- \mathrm{i} \sum_{j=1}^N \Gamma_j A_{ij}(\hat{\nabla}) 
 - \nu \hat{\nabla} \mathcal{I} \hat{\nabla}
 + \frac{\mathrm{i}}{2} \frac{\hat{\Gamma}_i}{\Gamma_i} K_i(\hat{\nabla}) \Biggr\} W_{ii} \eqcomma
\label{lagrangian}
\end{multline}
which one obtains immediately from the action \eqref{wirkung_ellipsen_d}. The action and thus the Lagrangian do not depend on the time derivative of the auxiliary field so the Euler-Lagrange equations for the variation with respect to $\hat{\omega}$ reduce to
 \begin{equation}
 \frac{\partial \mathcal{L}}{\partial \hat{f}_{i}} = 0
\label{eulers}
\end{equation}
with $\hat{f}_{i} \in [\hat{\Gamma}_i,\hat{\vx}_{i_k},\hat{C}_{i_{kl}}]$ for $k,l = (1,2)$.
\begin{subequations}
The variation with respect to $\hat{\Gamma}_i$ yields 
\begin{multline}
 0  = \Biggl\{ \frac{\dot{\Gamma}_i}{\Gamma_i} -  \dot{\vx}_i \cdot \hat{\nabla} + \halb \hat{\nabla} \dot{C}_i \hat{\nabla} 
 - \mathrm{i} \sum_{j=1}^N \Gamma_j A_{ij}(\hat{\nabla}) \\
 -\nu \hat{\nabla} \mathcal{I} \hat{\nabla} + \mathrm{i} \frac{\hat{\Gamma}_i}{\Gamma_i} K_i(\hat{\nabla}) \Biggr\}
  \, W_{ii} \eqcomma
\label{ib}
\end{multline}
the variation with respect to $\hat{\vx}_i$ leads to
\begin{multline}
 0  = \Biggl\{ \frac{\dot{\Gamma}_i}{\Gamma_i} -  \dot{\vx}_i \cdot \hat{\nabla} + \halb \hat{\nabla} \dot{C}_i \hat{\nabla} 
\\ - \mathrm{i} \sum_{j=1}^N \Gamma_j A_{ij}(\hat{\nabla})
 -\nu \hat{\nabla} \mathcal{I} \hat{\nabla} \Biggr\}
 \hat{\nabla}  \, W_{ii}
\label{iib}
\end{multline}
and the variation with respect to $\hat{C}_i$ provides
\begin{multline}
 0  = \Biggl\{ \frac{\dot{\Gamma}_i}{\Gamma_i} -  \dot{\vx}_i \cdot \hat{\nabla} + \halb \hat{\nabla} \dot{C}_i \hat{\nabla} 
 - \mathrm{i} \sum_{j=1}^N \Gamma_j A_{ij}(\hat{\nabla})
\\
 -\nu \hat{\nabla} \mathcal{I} \hat{\nabla} + \mathrm{i} \frac{\hat{\Gamma}_i}{\Gamma_i} K_i(\hat{\nabla}) \Biggr\}
 \hat{\nabla} \hat{\nabla}^T \, W_{ii} \eqdot
\label{iiib}
\end{multline}
\end{subequations}
We make a Gaussian ansatz for the correlation function $Q$ according to $Q(\vec{k}) =  q \exp[- \frac{1}{2} \vec{k} \Tilde{Q} \vec{k}]$. The width of $Q$ determines the forcing scale $\lambda$. 
We expand $A_{ij}(\hat{\nabla})$ and $K_i(\hat{\nabla})$ in equations \eqref{ib}-\eqref{iiib} according to
\begin{subequations}
 \begin{multline}
  A_{ij}(\hat{\nabla}) \approx \mathrm{i} \dintkk e^{- \mathrm{i} \vk' \cdot \vx - \halb \vk' (C_i+C_j) \vk'}
\\ \times \left[1 + \frac{\mathrm{i}}{2} \hat{\nabla} C_i \vk' + \frac{\mathrm{i}}{2} \vk' C_i \hat{\nabla} \right] \vu(\vk') \cdot \hat{\nabla}
 \end{multline}
and
\begin{equation}
 K_i(\hat{\nabla}) \approx \mathrm{i} q \left[1 - (\hat{\vx}_j-\vx_j) \cdot \hat{\nabla} + \halb \hat{\nabla} (\Tilde{Q} + \hat{C}_i - C_i) \hat{\nabla} \right] 
\end{equation}
\end{subequations}
and define ($\alpha_i^{(k)}$ and $\beta_i^{(k)}$ is of \hbox{$k$-th} order in $\hat{\nabla}$)
\begin{align*}
 \alpha_i^{(1)} & = \sum_{j=1}^N \Gamma_j \vec{U}_{ij}(\vx_i-\vx_j) \cdot \hat{\nabla} \eqcomma
\\
\alpha_i^{(2)} & = \hat{\nabla} \sum_{j=1}^N \Gamma_j \left[S_{ij}(\vx_i-\vx_j) C_i + C_i S_{ij}^T(\vx_i-\vx_j)\right] \hat{\nabla}
\\ & - \hat{\nabla} \nu \mathcal{I} \hat{\nabla} \eqcomma
\\
\beta_i^{(0)} & = - q \frac{\hat{\Gamma}_i}{\Gamma_i}   \eqcomma
\\
\beta_i^{(1)} & = q \frac{\hat{\Gamma}_i}{\Gamma_i} (\hat{\vx}_i - \vx_i) \cdot \hat{\nabla} \eqcomma
\\
\beta_i^{(2)} & = - q \frac{\hat{\Gamma}_i}{\Gamma_i} \halb \hat{\nabla}(\Tilde{Q} + \hat{C}_i - C_i) \hat{\nabla} \eqcomma
\end{align*}
where $\vec{U}_{ij}$ denotes the advection
\begin{equation}
 \vec{U}_{ij}(\vec{x})  =  \dintkk \vec{u}(\vec{k}') e^{-\mathrm{i} \vec{k}' \cdot \vec{x} -\frac{1}{2} \vec{k}' (C_i + C_j) \vec{k}'}
\end{equation}
and $S_{ij}$ denotes the deformation
\begin{equation}
S_{ij}(\vec{x})  =  -\mathrm{i} \dintkk \vec{u}(\vec{k}')  \vec{k}'^T  e^{-\mathrm{i} \vec{k}' \cdot \vec{x}-\frac{1}{2} \vec{k}' \left(C_i + C_j\right) \vec{k}'} \eqcomma
\end{equation}
which can be expressed as $\vec{U}_{ij}(\vec{x}) \overleftarrow{\nabla}^T$. We sort the obtained equations in orders of the $\hat{\nabla}$ operator and obtain with $\hat{\nabla}_{12} = \partial_{\hat{\vx}_{i_1}}+\partial_{\hat{\vx}_{i_2}}$
\begin{multline}
 \begin{pmatrix}
 1 & 0 & 0 \\
\hat{\nabla}_{12} & -1 & 0 \\
\hat{\nabla}_{12}^2 & - \hat{\nabla}_{12} & 1
\end{pmatrix}
\begin{pmatrix}
 \frac{\dot{\Gamma}_i}{\Gamma_i} \\
\dot{\vx}_i \cdot \hat{\nabla} \\
\halb \hat{\nabla} \dot{C}_i \hat{\nabla}
\end{pmatrix}
W_{ii}
\\ = 
\begin{pmatrix}
 -\beta^{(0)}_i \\
- \alpha^{(1)}_i - \beta^{(1)}_i  \\
- \alpha^{(2)}_i - \beta^{(2)}_i - \alpha^{(1)}_i \hat{\nabla}_{12} - \beta^{(0)}_i \hat{\nabla}_{12}^2
\end{pmatrix}
W_{ii} \eqdot
\label{matrixgln1}
\end{multline}
The matrix on the left side can be inverted and we obtain
\begin{equation}
   \begin{pmatrix}
 \frac{\dot{\Gamma}_i}{\Gamma_i} \\
\dot{\vx}_i \cdot \hat{\nabla} \\
\halb \hat{\nabla} \dot{C}_i \hat{\nabla}
\end{pmatrix}  
W_{ii}
\\
= 
\begin{pmatrix}
 - \beta^{(0)}_i \\
\alpha^{(1)}_i + \beta^{(1)}_i  \\
- \alpha^{(2)}_i - \beta^{(2)}_i 
\end{pmatrix}
W_{ii} \eqcomma
\label{matrixgln2}
\end{equation}
where we omitted terms proportional to $\hat{\nabla}_{12}$ on the right-hand side. From eq. \eqref{matrixgln2} we obtain the evolution equations
\begin{align}
\begin{split}
 \dot{\Gamma}_i &=  q \hat{\Gamma}_i \eqcomma \\
\dot{\vec{x}}_i & =  \sum_{j=1}^N \Gamma_j \vec{U}_{ij}(\vec{x}_i-\vec{x}_j)+ q \frac{\hat{\Gamma}_i}{\Gamma_i} (\vec{\hat{x}}_i - \vec{x}_i(t)) \eqcomma
\\
\dot{C}_i & =  \sum_{j=1}^N \Gamma_j \left[S_{ij}(\vec{x}_i-\vec{x}_j) C_i + C_i S_{ij}^{T} (\vec{x}_i - \vec{x}_j) \right] \\ &  + q \frac{\hat{\Gamma}_i}{\Gamma_i} \left(\Tilde{Q} + \hat{C}_i - C_i\right) + 2 \nu \mathcal{I} 
\end{split}
\label{instanton1}
\end{align}
for the most probable evolution of elliptic vortices.

An analogue calculation leads to the evolution equations for the elliptic structures in the auxiliary field
\begin{align}
\begin{split}
  \dot{\hat{\Gamma}}_i  = & 0 \eqcomma \\
  \dot{\hat{\vec{x}}}_i  = &  \sum_{j=1}^N \Gamma_j \vec{\hat{U}}_{ij}(\hat{\vec{x}}_i - \vec{x}_j) \eqcomma \\
\dot{\hat{C}}_i  = &  \sum_{j=1}^N \Gamma_j \left[ \hat{S}_{ij}\left(\hat{\vec{x}}_i - \vec{x}_j\right) \hat{C}_i + \hat{C}_i \hat{S}_{ij}^{T}\left(\hat{\vec{x}}_i - \vec{x}_j\right) \right] \\ & - 2 \nu \mathcal{I} \eqcomma
\end{split}
\label{instanton2}
\end{align}
where we defined the advection and deformation for the auxiliary field
\begin{align}
\begin{split}
 \vec{\hat{U}}_{ij}(\vec{x}) & =  \dintkk \vec{u}(\vec{k}') e^{- \mathrm{i} \vec{k}' \cdot \vec{x}-\frac{1}{2} \vec{k}' \left( \hat{C}_i + C_j \right) \vec{k}'} \eqcomma \\
\hat{S}_{ij}(\vec{x}) & =  \vec{\hat{U}}_{ij}(\vec{x}) \overleftarrow{\nabla}^T \eqdot
\end{split}
\end{align}
To derive these equations, one has to take into account the variation of the Lagrangian with respect to $\Gamma_i$, $\vx_i$ and $C_i$ as well as their time derivatives $\dot{\Gamma}_i$, $\dot{\vx}_i$ and $\dot{C}_i$. Note, that we omitted higher order terms in $\hat{C}_i$.

\section{Transition probability}
\label{section_transition_probability}
With the Instanton equation for the vorticity field $\omega(\vx,t)$ from equation \eqref{instantion_equations} we obtain the extremal action \hbox{$S_{\mathrm{e}} = - \frac{\mathrm{i}}{2} \dintkt \hat{\omega}(-\vec{k},t) Q(\vec{k}) \hat{\omega}(\vec{k},t)$} where the $k$ integration is Gaussian and leads to
% \begin{equation*}
%  S_{\mathrm{e}} = i \pi \int_{t^*}^0 \! \mathrm{d}t \, \left[ \hat{\Gamma}_1^2 \det (\Tilde{Q} + 2 \hat{C}_1)^{-\frac{1}{2}} + \hat{\Gamma}_2^2 \det (\Tilde{Q} + 2 \hat{C}_2)^{-\frac{1}{2}} \right]  \eqdot
% \label{s_ext_exp}
% \end{equation*}
\begin{equation*}
 S_{\mathrm{e}} = \mathrm{i} \pi \int_{t^*}^0 \! \mathrm{d}t \, \left[ \frac{\hat{\Gamma}_1^2}{\sqrt{ \det (\Tilde{Q} + 2 \hat{C}_1)}} + \frac{\hat{\Gamma}_2^2}{\sqrt{ \det (\Tilde{Q} + 2 \hat{C}_2)}} \right]  \eqdot
\label{s_ext_exp}
\end{equation*}
Here we omitted the overlap in accordance with our assumptions. In the Instanton approximation $\mathcal{Z} \approx e^{\mathrm{i} S_{\mathrm{e}}}$ we get an zeroth-order approximation to the transition probability density 
\begin{equation}
 f_C\left(  \left. \{\vx_i^0,\Gamma_i^0\}\right|\{\vx_i,\Gamma_i\}\right)  \approx e^{\mathrm{i} S_{\mathrm{e}} }
\end{equation}
which depends on the shape $C$ of the vortices at time $t^*$. The time $t^*$ has to be extracted from \hbox{$\Gamma_i(t) = q \hat{\Gamma}_i (t-t^*) + \Gamma_i(t^*)$}. The transition probability $f$ can be written as
\begin{multline}
  f\left(  \left. \{\vx_i^0,\Gamma_i^0\}\right|\{\vx_i,\Gamma_i\}\right)  =
  \\ \int \! \mathrm{d}C \, p(C) f_C\left(  \left. \{\vx_i^0,\Gamma_i^0\}\right|\{\vx_i,\Gamma_i\}\right)
\label{transition_probability}
\end{multline}
where $p(C)$ is a distribution of initial shapes $C$.
The investigation of this expression is left for future work.

\bibliographystyle{unsrt}
\bibliography{my}

\end{document}